\documentclass[aps,prl,showpacs,twocolumn,superscriptaddress]{revtex4}

\usepackage{amsmath}
\usepackage{amstext}
\usepackage{amssymb}
\usepackage{graphicx}
\usepackage{bbm,multirow}

\def\tr{\,{\rm tr}\,}

\def\ave#1{\langle #1 \rangle}
\def\ii{{\rm i}}
\def\vec#1{\underline{#1}}

\def\tit#1{}
\def\etal#1{ {\em et al.}}

\begin{document}

\title{Thermalization and ergodicity in one-dimensional 
many-body open quantum systems}

\author{Marko \v Znidari\v c}
\affiliation{Department of Physics, Faculty of Mathematics and Physics,
  University of Ljubljana, Ljubljana, Slovenia}

\author{Toma\v z Prosen}
\affiliation{Department of Physics, Faculty of Mathematics and Physics,
  University of Ljubljana, Ljubljana, Slovenia}

\author{Giuliano Benenti}
\affiliation{CNISM, CNR-INFM, and Center for Nonlinear and Complex Systems,
  Universit\`a degli Studi dell'Insubria, Via Valleggio 11, 22100
  Como, Italy}
\affiliation{Istituto Nazionale di Fisica Nucleare, Sezione di Milano,
  Via Celoria 16, 20133 Milano, Italy}

\author{Giulio Casati}
\affiliation{CNISM, CNR-INFM, and Center for Nonlinear and Complex Systems,
  Universit\`a degli Studi dell'Insubria, Via Valleggio 11, 22100
  Como, Italy}
\affiliation{Istituto Nazionale di Fisica Nucleare, Sezione di Milano,
  Via Celoria 16, 20133 Milano, Italy}
\affiliation{Centre for Quantum Technologies,
  National University of Singapore, Singapore 117543}

\author{Davide Rossini}
\affiliation{International School for Advanced Studies (SISSA),
  Via Beirut 2-4, I-34014 Trieste, Italy}

\date{\today}

\begin{abstract}
Using an approach based on the time-dependent density-matrix renormalization
group method, we study thermalization in spin chains locally coupled
to an external bath.
Our results provide evidence that quantum chaotic systems do thermalize,
that is, they exhibit relaxation to an invariant ergodic state which, 
in the bulk, is well approximated by the grand canonical state.
Moreover, the resulting ergodic state in the bulk does not depend
on the details of the baths. On the other hand, for integrable systems
we found that the invariant state in general depends on the bath and is 
different from the grand canonical state.
\end{abstract}

\pacs{03.65.Yz, 75.10.Pq, 05.45.Mt}



\maketitle

The emergence of canonical ensembles in quantum statistical mechanics
from first principles is one of the key remaining old questions
of theoretical physics.
Even the definition of the temperature at the nano-scale poses
a challenge~\cite{Hartmann:04}.
Namely, the main question is how to ``derive'' the canonical distribution?
It has been realized that the canonical distribution is in a way ``typical'':
provided the overall system describing the environment 
plus a central system is
in a generic pure state, the reduced state of the central system is with
high probability canonical~\cite{Gemmer:03}.
However, how precisely
the canonical distribution arises from dynamical laws,
without a priori statistical assumptions, is still unclear.
Motivation in the study of this fundamental aspect of nonequilibrium
physics also comes
from some recent experiments with ultracold bosonic gases,
where absence of thermalization in closed, integrable,
strongly correlated quantum systems
has been observed~\cite{kinoshita06}.

For closed many-body systems, integrability is believed to play
a crucial role in the relaxation to the {\em Steady State} (SS):
the nonequilibrium dynamics of a chaotic system
is expected to thermalize
at the level of individual eigenstates~\cite{ETH},
as numerically observed in several physical models~\cite{closed_nonint}.
By contrast, for systems with non trivial integrals of motion,
SSs usually carry memory of the initial conditions
and are not canonical:  maximizing the entropy while keeping
the values of constants of motion fixed results in a generalized Gibbs
ensemble~\cite{Rigol}.
Much less is known about the relaxation to the SS
for {\em open quantum systems}~\cite{henrich:05}; 
this is what we are going to address in this paper.
We provide numerical evidence that, analogously to closed
systems, the occurrence of thermalization
is strictly related to system's integrability,
irrespective of the fine details of the baths.
In particular we show that {\em locally} coupling a quantum chaotic
many-body system to an environment is enough for a SS 
of the central system to be very close, in the bulk, 
to the canonical or grand canonical state (GCS). 
On the contrary, if the system is integrable, the constants
of motion in general prevent thermalization and the form of the SS 
sensitively depends on the bath coupling operators.
We show that the numerical description of an open quantum system in terms
of a Lindblad equation with {\em local} coupling to the reservoirs is in 
some sense a {\em computationally efficient, minimal model of thermalization}.
Such result paves the way for future simulations of 
quantum transport in large many-body quantum systems.

The time evolution for a generic state $\rho$ of an open quantum system
can be described, under certain approximations, by a Lindblad
master equation~\cite{breuer:BOOK}:
\begin{equation}
  \frac{{\rm d}}{{\rm d}t}{\rho} = \frac{\ii}{\hbar} [ \rho, {\cal H} ]
  + {\hat {\cal L}}_{\rm B}\rho,
  \label{eq:Lin}
\end{equation}
where ${\cal H}$ is the Hamiltonian of the autonomous system,
while the dissipation
$\hat{\cal L}_{\rm B} = \gamma \sum_k \left( [ L_k \rho,L_k^{\dagger} ] +
[ L_k,\rho L_k^{\dagger} ] \right)$ is parametrized
by certain Lindblad operators $L_k$
(hereafter we set $\hbar = k_B = 1$ and, unless noted otherwise, $\gamma=1$).
The derivation of Eq.~\eqref{eq:Lin}
from first principles, i.e., from the Hamiltonian evolution
of a system plus environment is rather tricky~\cite{breuer:BOOK};
however it is the most general form of a completely positive,
trace preserving, dynamical semi-group.
Taking it for granted, we ask ourselves if, within this approximation,
a {\em finite} many-body system can thermalize when coupled
via some Lindblad operators $L_k$
acting only {\em locally} just on {\em few} degrees of freedom.

To elucidate the role covered
by chaoticity in the thermalization process,
we consider
prototype one-dimensional spin-$1/2$-chain models
with nearest neighbor interactions:
${\cal H} = \sum_{l=0}^{n-2} h_{l,l+1}$
($h_{l,l+1}$ denoting the local energy density,
and $n$ being the chain length).
As we shall see, the chosen 
models exhibit a crossover from integrable to
chaotic regime when a suitable parameter in their 
Hamiltonians is varied. 
With the term ``chaotic'' we refer, as usual, to a
system whose bulk energy spectrum of highly excited
levels obeys a random matrix statistics~\cite{haake}; 
in particular, the level spacing statistics (LSS) $p(s)$ 
is well approximated by the Wigner-Dyson distribution 
$p_{\rm WD}(s)$~\cite{haake}, whereas in 
an integrable system LSS typically turns out
to be Poissonian, $p_{\rm P}(s)$.

We assume local coupling to the reservoirs, i.e.,
the dissipator $\hat{\cal L}_{\rm B}$ acts only on the $m$ ($\ll n$) 
leftmost $(l)$ and rightmost $(r)$ spins:
$\hat{\cal L}_{\rm B}=
\hat{\cal L}^{l}_{\rm B}\otimes \hat{\openone}_{\rm bulk}
\otimes \hat{\cal L}^{r}_{\rm B}$. 
We construct $\hat{\cal L}_{\rm B}$ by generalizing the
method discussed in Ref.~\cite{JSTAT:09}. 
For this purpose, 
we first consider the GCS for the spin chain,
\begin{equation}
  \rho_{\cal G} (T,\mu) = Z^{-1}
  \exp\left[-({\cal H} - \mu\, \Sigma^{\rm z})/T \right],
  \label{eq:grandc}
\end{equation}
where
$\Sigma^{\rm z}=\sum_{l=0}^{n-1} \sigma_l^{\rm z}$ is the total magnetization
[$\sigma^\alpha_j$ ($\alpha={\rm x},{\rm y},{\rm z}$)
being the Pauli operators for the $j$th spin],
$T$ the temperature, $\mu$ the ``chemical potential'', and
$Z = \tr{\left[\exp{(-({\cal H} - \mu\,\Sigma^{\rm z})/ T)}\right]}$
the partition function.
Given a target temperature $T_{\rm targ}$ and a chemical potential
$\mu_{\rm targ}$, the reduced $m$-spin target density matrix 
$\rho^{\lambda}_{\rm targ}$, $\lambda \in \{l,r\}$, is obtained after tracing
$\rho_{\cal G} (T_{\rm targ},\mu_{\rm targ})$ over all but
the $m$ leftmost/rightmost spins.  
We finally require that 
$\rho^{\lambda}_{\rm targ}$ is the unique eigenvector of
$\hat{\cal L}^{\lambda}_{\rm B}$ with eigenvalue $0$, while 
all other eigenvalues are equal to $-1$. 
Such a choice produces, in absence of ${\cal H}$ and 
for a given spectral norm of $\hat{\cal L}^{\lambda}_{\rm B}$, 
the {\em fastest} convergence to $\rho^{\lambda}_{\rm targ}$~\cite{lindbladnote}.
In the presence of ${\cal H}$ we obtain, for up to $n\approx 100$ spins, the SS 
solution of Eq.~\eqref{eq:Lin}
numerically by using a time-dependent Density Matrix Remormalization
Group (tDMRG) method with a Matrix Product Operator (MPO) ansatz~\cite{tDMRGreview}.

In the following we are interested in the asymptotic state reached,
independently of initial conditions, after a
long time, $\rho_{\rm SS} \equiv \lim_{t \to \infty}\rho(t)$. 
In all simulations we carefully checked that the simulation time
was long enough to reach convergence, which is exponential.
Since Lindblad operators act only locally and $\rho_{\cal G}(T,\mu)$
is invariant for the unitary part of Eq.~\eqref{eq:Lin}, $\rho_{\rm SS}$
cannot be equal to the GCS,
unless it is also an eigenstate of the dissipator $\hat{\cal L}_{\rm B}$.
In other words, one can have $\rho_{\rm SS}=\rho_{\cal G}(T,\mu)$ only if
$\rho_{\cal G}(T,\mu)=\rho^{l}_{\rm targ}\otimes \rho_{\rm bulk}\otimes
\rho^{r}_{\rm targ}$, i.e., if the GCS
is separable with respect to the border $m$ spins which are used
in the coupling. Nevertheless for chaotic systems, as we shall see,
sufficiently far from the boundaries the state
is arbitrarily close to $\rho_{\cal G}(T,\mu)$, regardless of the 
entanglement with the coupled parts.

\begin{figure}[!t]
  \centerline{\includegraphics[angle=-90,scale=0.35]{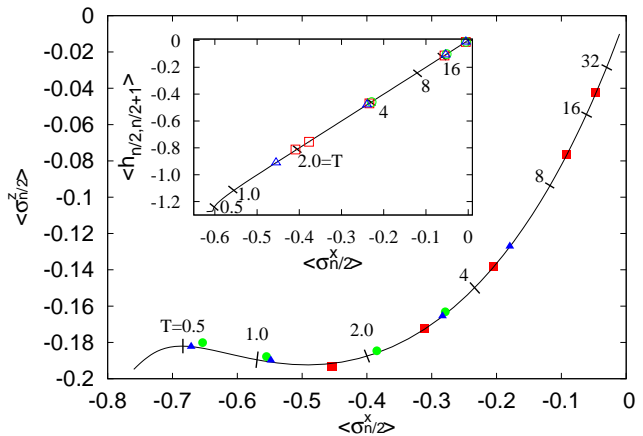}}
  \caption{(Color online). One-spin observables for the SS (symbols)
    agree with the theoretical canonical ones (full curve) to less than
    $0.5\,\%$ for the chaotic Ising model (main plot).
    For the integrable model (inset) a comparable agreement is observed
    by looking at energy density (since $\ave{\sigma^z_{n/2}}\equiv 0$)
    vs.~$\ave{\sigma^{\rm x}_{n/2}}$.
    Squares are for $n=16$ and uniform couplings;
    circles (triangles) for $n=16$ ($n=40$) and couplings 
    $J_l$ switched on over a layer of
    thickness $\tau = 4$, with $\gamma = 0.2$.
    Marks on theoretical curves show the temperature.}
  \label{fig:xzCcti}
\end{figure}

Let us start our numerical investigations by 
considering a spin-$1/2$ Ising chain
in a tilted magnetic field, described by the energy density
\begin{equation}
  h_{l,l+1} = J_l \sigma_l^{\rm z} \sigma_{l+1}^{\rm z} +
  \frac{b_{\rm x}}{2}(\sigma_{l}^{\rm x}+\sigma_{l+1}^{\rm x}) +
  \frac{b_{\rm z}}{2}(\sigma_{l}^{\rm z}+\sigma_{l+1}^{\rm z}).
  \label{eq:ising}
\end{equation}
Its only conserved quantity is the total energy, therefore the
expected invariant state is the canonical one $\rho_{\cal G}(T,0)$.
To check thermalization, we solved the master equation
for two different sets of parameters:
(i) a transverse field $b_{\rm x}=1,b_{\rm z}=0$,
for which the model is integrable and exhibits a Poissonian LSS;
(ii) a tilted field $b_{\rm x}=1,b_{\rm z}=1$, for which it is chaotic
with a Wigner-Dyson LSS~\cite{PRE}
(if not specified, we take
$J_l = 1$ and couple two border spins, $m=2$).
With the obtained $\rho_{\rm SS}$, we evaluated expectation
values of several one- and two-spin observables in the bulk of the chain,
and compared them to the theoretical ones as given
by the canonical state $\rho_{\cal G}(T,0)$.

In the main plot of 
Fig.~\ref{fig:xzCcti} we show one-spin expectation values
$\ave{\sigma^{\alpha}_{n/2}}=\tr{(\rho_{\rm SS} \, \sigma_{n/2}^{\alpha})}$
for the chaotic case:
all numerical points fall on the curve given by theoretical
expectation values for a canonical state.
The same happens in the integrable
Ising model. Such irrelevance of integrability is a peculiarity of certain
few-body observables, similarly
to what observed in a different context of out-of-equilibrium
dynamics in closed systems~\cite{rossini09}.
Quite remarkably,
we could not reach temperatures in the bulk below $\approx 1.7$
(see squares in Fig.~\ref{fig:xzCcti}),
even by using very small $T_{\rm targ}\approx 0$.
The reason resides in the already mentioned boundary effects due to
entanglement between the boundary two spins and the bulk chain,
which makes the cooling difficult.
This must be contrasted with a zero attainable temperature
in the case of separable states~\cite{Giovannetti}.
For entangled states though, our results show that
to lower the minimal attainable temperature one has to reduce the effect
of interaction at the boundaries which is responsible for entanglement.
One way to do this is by switching on the interaction gently over
a boundary layer of certain thickness $\tau$,
$J_l=\sin{(\frac{l}{\tau}\frac{\pi}{2})}$ 
($J_{n-2-l}=\sin{(\frac{l}{\tau}\frac{\pi}{2})}$), 
for $l=0,\ldots,\tau-1$, at the left (right) end
and using a weaker coupling $\gamma$
(circles and triangles in Fig.~\ref{fig:xzCcti}).

\begin{figure}[!t]
  \centerline{\includegraphics[angle=-90,scale=0.33]{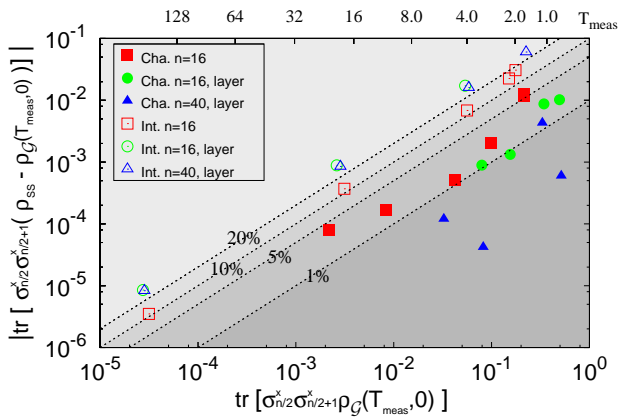}}
  \caption{(Color online). Absolute differences
    in the expectation value of a two-spin observable
    $\sigma_{n/2}^{\rm x}\sigma_{n/2+1}^{\rm x}$ between the SS
    and the theoretical canonical state, in the case of
    chaotic (full symbols) and integrable Ising model (empty symbols).
    Dashed lines denote constant relative error.}
  \label{fig:Dxx}
\end{figure}

To make comparison between $\rho_{\rm SS}$ and $\rho_{\cal G}(T,0)$
quantitative, we determined the ``measured'' temperature $T_{\rm meas}$
to which $\rho_{\rm SS}$ corresponds, which is in general different
from $T_{\rm targ}$, due to boundary effects.
Assuming that the SS is canonical in the bulk, one can extract $T_{\rm meas}$
by comparing observables that uniquely set the temperature.
For Ising model~\eqref{eq:ising}, the energy density is sufficient,
therefore we used the condition
$\tr{[h_{n/2-1,n/2} \, \rho_{\rm SS}]} \equiv
 \tr{[h_{n/2-1,n/2} \, \rho_{\cal G}(T_{\rm meas},0)]}$
to compute $T_{\rm meas}$.
We then calculated theoretical expectation values of other observables,
through $\rho_{\cal G}(T_{\rm meas},0)$; a comparison
with the corresponding values for the reached SS
may serve as an indicator of the {\em quality of thermalization}.
In Fig.~\ref{fig:Dxx} we show differences between expectation
values of $\sigma_l^{\rm x} \sigma_{l+1}^{\rm x}$,
computed with $\rho_{\rm SS}$ and $\rho_{\cal G}(T_{\rm meas},0)$,
for both chaotic and integrable Ising chains.
A marked distinction between the two cases appears.
First, in the chaotic model errors are much smaller than in the
integrable one; second, switching $J_l$ gradually, which should
decrease errors due to smaller boundary effects, in the integrable case
even worsens the situation. The integrable Ising model therefore
does not relax to a canonical state in the bulk.
Similar results are obtained for other few-spin observables,
as well as for the lowest moments of the energy distribution:
we evaluated  $\ave{[({\cal H}_6 - \ave{{\cal H}_6})/5]^p}$
($p=2,...,5$ and ${\cal H}_6$ is the Hamiltonian of the 6 central spins) 
on the states $\rho_{\rm SS}$ and $\rho_{\cal G}(T_{\rm meas},0)$.
In a chain of $n=40$ spins relative errors are never greater
than $1 \%$ in the chaotic case, and are typically
an order of magnitude larger in the integrable case.

\begin{figure}[!t]
  \begin{center}
    \includegraphics[scale=0.41]{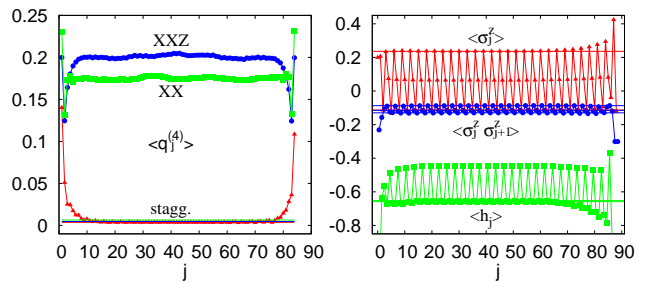}
    \caption{(Color online). SS (symbols) and GCS (full lines) expectation 
      values of $q_j^{(4)}$ (left panel), 
      $\sigma_j^{\rm z}$, $\sigma_j^{\rm z}\sigma_{j+1}^{\rm z}$, and 
      $h_j$ (right panel) for the Heisenberg model
      with $n=89$ spins,
      $T_{\rm targ}=4$, $q_{\rm targ}=2$, 
      $m=3$, $J_l=1$.
      In the left panel, 
      ``XX'' and ``XXZ'' refer to two integrable cases
      without magnetic field (respectively at $\Delta = 0, \, 0.5$),
      ``stagg.'' to the chaotic case with $\Delta=0.5$ and period-3
      staggered field with $B=2$.
      The curves in the right panel are for the chaotic case only.
      The GCS $\rho_{\cal G}(T_{\rm meas}=5.851,\mu_{\rm meas}=-0.534)$ for the 
      chaotic case is obtained
      by matching $\ave{h_{3l+1,3l+2}}$ and $\ave{\sigma^{\rm z}_{3l+1}}$
      for which lines are not shown in the right panel.
    }
    \label{fig:XXZstagg2}
  \end{center}
\end{figure}

To corroborate the importance of system's integrability on the
convergence to invariant statistical ensembles, we consider another
prototype model of interacting spins: the Heisenberg XXZ chain
in a magnetic field, described by the energy density
\begin{equation}
  h_{l,l+1} = J_l (\sigma_l^{\rm x} \sigma_{l+1}^{\rm x} +
  \sigma_l^{\rm y} \sigma_{l+1}^{\rm y}+\Delta \sigma_l^{\rm z}
  \sigma_{l+1}^{\rm z})+\frac{b_{l}}{2}\sigma_{l}^{\rm z} +
  \frac{b_{l+1}}{2}\sigma_{l+1}^{\rm z}.
  \label{eq:heis}
\end{equation}
If the field is homogeneous the model is integrable and possesses,
besides energy and magnetization, an infinite sequence of conserved
quantities~\cite{Grabowski}. On the other hand, integrability can be 
broken, e.g., simply by means of a period-3
staggered magnetic field, $b_{3k}=-B, \, b_{3k+1}=-B/2, \, b_{3k+2}=0$. 
In order to highlight the lack of thermalization in the integrable
regime $B=0$, we target a non-Gibbsian state different
from the GCS;
namely, we use 
$\rho_{\rm non-{\cal G}}(T,q) \sim \exp{(-{\cal H}/T+q\, Q_4)}$,
with $Q_4=-h_{0,1}-h_{n-2,n-1}+\sum_{l=0}^{n-4} q^{(4)}_l$, $q^{(4)}_l =
\sigma_l^{\rm x} \sigma_{l+1}^{\rm z} \sigma_{l+2}^{\rm z}\sigma_{l+3}^{\rm x} +
\sigma_l^{\rm y} \sigma_{l+1}^{\rm z} \sigma_{l+2}^{\rm z}\sigma_{l+3}^{\rm y}$,
being a conserved charge for an open chain with $\Delta=0$
and $b_l=0$~\cite{Grabowski}.
The idea is that, using a $q_{\rm targ}\ne 0$, in the integrable regime 
the SS exhibits strong deviations from the GCS, corresponding to $q=0$,
while we expect chaotic dynamics to drive the bulk towards the GCS.
Such expectation is confirmed by our numerical data. 
In Fig.~\ref{fig:XXZstagg2} we show the spatial dependence of various
observables for integrable, as well as for chaotic cases.
In the integrable cases deviations from the GCS expectations are large,
while they become very small for a chaotic system.
Analogously to the Ising model, we checked this statement
also for other few-spin observables
(we found that, in presence of chaos, the largest discrepancy among
all the one- and two-spin observables amounts to $2\times 10^{-4}$);
layered interactions in the integrable model do not
help in thermalizing the system.

\begin{figure}[!t]
  \begin{center}
    \includegraphics[scale=0.3]{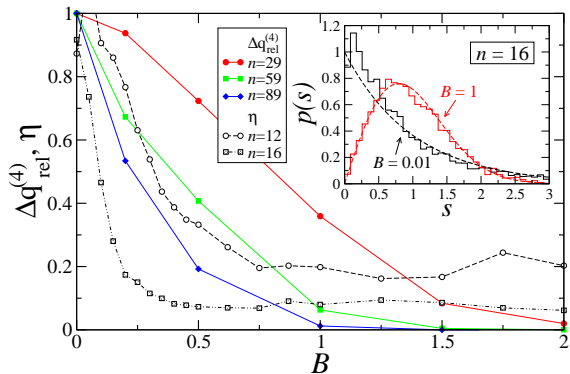}
    \caption{(Color online).
      Relative differences
      $\Delta q^{(4)}_{\rm rel} = \Delta q^{(4)}(B)/\Delta q^{(4)}(0)$
      in $q^{(4)}$ expectation values on the SS
      and the GCS evaluated in the bulk of the Heisenberg model
      with $\Delta = 0.5$, as the staggering strength $B$
      is varied (full curves). Also shown is dependence of the $\eta$ function
      (dashed curves), characterizing the integrable-chaotic crossover.
      For both quantities the crossover takes place at smaller $B$ 
      with increasing $n$.
      Inset: two examples of LSS in the integrable ($B = 0.01$)
      and chaotic ($B=1$) regimes; dashed lines
      denote Poissonian and Wigner-Dyson statistics~\cite{haake},
      respectively.}
    \label{fig:transition}
  \end{center}
\end{figure}

A further confirmation of the role of integrability 
comes from a direct analysis of the quality of thermalization after gradually
switching on the perturbation that drives the crossover from 
integrability to chaos: the longitudinal field $b_z$
in Eq.~\eqref{eq:ising} or the staggering intensity $B$ in Eq.~\eqref{eq:heis}.
As shown in Fig.~\ref{fig:transition} for the Heisenberg model, 
such crossover is conveniently detected by the parameter $\eta\equiv 
\int{|p(s)-p_{\rm WD}(s)|{\rm d}s}/\int{|p_{\rm P}(s)-p_{\rm WD}(s)|{\rm d}s}$;
$\eta=1$ and $\eta=0$ correspond to Poissonian 
and Wigner-Dyson distributions, respectively.
In the same figure we also plot deviations in $\ave{q^{(4)}}$
evaluated on the SS and on the corresponding GCS:
$\Delta q^{(4)}=\tr{ [ q^{(4)}_l (\rho_{\rm SS}-\rho_{\cal G}
(T_{\rm meas},\mu_{\rm meas}))]}$
as the strength $B$ of the staggered magnetic field is increased.
The progressive onset of chaos gradually improves the quality of 
thermalization, being $\Delta q^{(4)}$ a monotonic decreasing function of $B$.
Moreover, the strength of the staggered field required to converge to 
the GC expectation value drops with the system size.

In conclusion, we have shown that, within the Lindblad equation formalism,
coupling a one-dimensional quantum chaotic system locally to a bath results
in a SS being equal to the invariant (grand)canonical state, far away from
the coupled sites. In contrast, integrable systems do not thermalize and their
SSs exhibit strong deviations from the (grand)canonical state,
depending on the details of the coupling.
The fact that for chaotic systems the SS does not depend on the details
of the coupling, shows that very likely the same result would be obtained
even for a harder-to-treat Hamiltonian evolution of a system plus environment
or for higher dimensional systems.
Our method should be applicable also to non-equilibrium situations.
Indeed, by locally coupling a system to two or several baths at different 
values of temperature and chemical potentials, one should be able 
to efficiently control local thermalization.
Thus, our results might open significant new perspectives in the simulation
of quantum transport in many-body quantum systems in contact with thermal and
chemical baths.

We thank V. Giovannetti for useful discussions.
M\v Z and TP are supported by the Program P1-0044, and the Grant  J1-2208, 
of the Slovenian Research Agency.


\begin{thebibliography}{100}
\bibitem{lindbladnote}
 To construct such $\hat{\cal L}^{l}_{\rm B}$ (and 
 similarly for $\hat{\cal L}^{r}_{\rm B}$),
 we first diagonalize $\rho^{l}_{\rm targ}$, thus getting eigenvalues 
 $d_i$ and eigenvectors, $\rho^{l}_{\rm targ}=V^\dagger d V$. 
 Then we obtain $\hat{\cal L}^{l}_{\rm B}$ after rotating 
 the ``diagonal'' Lindblad superoperator $\hat{\cal L}^{\rm diag}_{\rm B}$,
 $\hat{\cal L}^{l}_{\rm B}=R \hat{\cal L}^{\rm diag}_{\rm B} R^{\rm T}$,
 with the orthogonal matrix
 $R_{\vec{\alpha},\vec{\beta}}=\tr{(V^\dagger \sigma^{\vec{\alpha}}\,
 V \sigma^{\vec{\beta}})}/4$ 
 ($\sigma^{\vec{\alpha}}=\sigma^{{\alpha_1}}\otimes
 \cdots \otimes \sigma^{{\alpha_m}}$). 
 A set of $4^m$ Lindblad operators leading to the diagonal 
 superoperator $\hat{\cal L}^{\rm diag}_{\rm B}$ is given by
$
  L_{i_1 i_2 \ldots i_m} = \sqrt{\frac{d_k}{2\cdot 4^m}}\, r^{i_1}
  \otimes r^{i_2} \otimes \cdots \otimes r^{i_m},
  \label{eq:L}
$
 with $i_j \in \{0,1,2,3\}$,
 $k=\sum_{j=1}^m{ (i_j\, \hbox{mod}\, 2 )\cdot 2^{j-1}}$, and
 $r^0 = \sigma^{\rm x}+\ii \sigma^{\rm y},r^1=\sigma^{\rm x} -
 \ii \sigma^{\rm y},r^2=\mathbbm{1}+\sigma^{\rm z},r^3 =
 \mathbbm{1}-\sigma^{\rm z}$.
\end{thebibliography}
\end{document}